# The decreasing level of Toshka Lakes seen from space

Amelia Carolina Sparavigna
Dipartimento di Fisica,
Politecnico di Torino, Torino, Italy

Toshka Lakes are lakes recently formed in the Sahara Desert of Egypt, by the water of the Nile, conveyed from the Nasser Lake through a canal in the Toshka Depression. From space, astronauts noticed the growing of a first lake, the easternmost one, in 1998. Then additional lakes grew in succession due west, the westernmost one between 2000 and 2001. In fact, sources of precious information on Toshka Lakes are the pictures takes by the crews of space missions and the satellite imagery. They show that, from 2006, the lakes started shrinking. A set of recent images displays that the surface of the easternmost lake is strongly reduced.

Toshka Lakes is the name given to the lakes, quite recently formed, in the Sahara Desert of Egypt. These new lakes are endorheic, that is, the water can never flow out this region toward the sea. They are near the Nasser Lake (see the map in Fig.1), the huge lake created by the Aswan High Dam, built in 1964-1968. As Wikipedia is reporting [1], in 1978, Egypt began to build the Sadat Canal, originating from the Lake Nasser and passing through the Wadi Toshka, to allow the water, when it is overflowing a specific level, to be drained off to flood a depression located at the south end of a desert limestone plateau. This is the Toshka Depression. In fact, the Egyptian government decided to undertake a development project of the surrounding region, known as the "New Valley Project". The project intends to extend this waterway till the Kharga oasis.

It was in late 1990s [1] that the water started flowing in the canal. This waterway, feeding the depression with the water from Nile, was prepared in order to reduce the water losses. The water, in order to flow the Sadat Canal needs a pumping station, which is situated north of Abu Simbel. From space, astronauts noticed the growing of a first lake, the easternmost one, in November 1998 [1]. They saw the other lakes grown in succession westward, the westernmost lake between September 2000 and March 2001. A public source of information on Toshka Lakes is in fact in the pictures taken by the crews of space missions. Starting from 2006, the images show that the lakes started shrinking, because of a reduction of water supply [3]. One of the lakes dried out.

As reported in [3], the flooding of the Toshka Depression had created the four wide lakes, with their maximum surface area in 2002, of ca 1,500 km$^2$, having about $5 \times 10^9$ m$^3$ of water. Some of the water evaporated and some recharges the underlying aquifer [3]. At the beginning of the flooding, the fauna and flora of Nile invaded the Toshka valley [3], with a resultant very rich population of fishes. But, unfortunately, the water supply reduced. The western-most lake shows a rapidly increasing salinity. As the researches are observing, the zooplankton reacts to the increase of salinity by a striking impoverishment [3]. The life in the lake started to decline.

A publication [4] estimated the loss of water from the lakes using remote sensing and GIS, collecting a set of images and processing them to show the aerial extend shrinkage of these lakes from 2002 up to 2007. By means of a spatial analysis of bathymetry in a digital elevation model, the authors found that the loss rate is around 2.5 m/year. They noted that the lakes stored around 25.26 billion cubic meters of water in 2002, but in 2006 the stored water was reduced to 12.57 billion cubic. The authors conclude that the location of these lakes in such hyper-arid zone is responsible for a high loss rate, mainly caused by the evaporation, because they found the percolation of water through the ground quite limited. They say that "It is strongly recommended that measures must be taken to maximize the benefits of these huge, exceptional water resources before totally lost via evaporation. Additionally, geo-environmental problems are likely to arise from the concentration of salts when the lakes dry up".

Let us see what happened by means of the NASA satellite imagery. The "NASA Earth Observatory" allow us to see the evolution of the western Toshka Lakes (see Fig.2). In the upper image, we have a figure of the lakes as soon as they formed, 2001. But in 2005, the water level was strongly reduced. In Fig.3, the "NASA Earth Observatory" dated 2005 is compared with an image that we can obtain (on date July 21, 2011) from Google Maps. It is not possible for the author to tell when the Google image had been recorded. But, surely it is more recent, because we see a further reduction of the surfaces of the lakes.

Let us note that reference [4] is dated 2007. But what is the current situation of the lakes?

Let us try to answer using NASA images again, and in particular, those of the easternmost Toshka Lake. In 2005, it was as shown in upper panel of Fig.4, as provided by the "Gateway to Astronaut Photography of Earth". The original image has been enhanced and rotated to compare it with that provided the Google Maps, displayed in the lower panel of Fig.4. As told for Fig.3, it is not possible for me to date the Google image; surely after 2005, because comparing the images, we note that the surface of the lake had shrunk.

But the "Gateway to Astronaut Photography of Earth" has a set of images of March 31, 2011, quite recent then, that allowed me to create a composite map of the current aspect of the lake. The map is proposed in Fig.5. This map had been rotated and enhanced as discussed in Ref. 5 (the image processing proposed in [5] has interesting applications in archaeology, see [6]). The processed images have so many details that we can use them to see the evolution of the lake. This is proposed in Fig.6, where we see a part of the lake, as it evolved from 2005 to March 31, 2011. In this figure we pass from the image dated 2005, to that of 2011, through that from Google Maps, of unknown date. This part of the easternmost lake was large in 2005, shrinking between 2005 and 2011, and reduced to one half the original in 2011, as it can be seen in Figure 7, where the superposition of the two NASA images, of 2005 and 2011, is proposed. We see the surface of the lake strongly decreased.

Let me continue with a little bit of history. Egypt had already lived a similar situation, but on a longer period and more than two thousand years ago. In Egypt there are several lakes; one is the Lake Moeris, an ancient saltwater lake in the northwest of the Faiyum Oasis. During the prehistory, the lake had freshwaters due to the high flood of Nile. From the XII Dynasty, in 2300 BC, the natural waterway from the river to the lake was widened and deepened to make a canal which is now known as the Bahr Yussef, "the waterway of Joseph" [7]. This canal served for the purpose to control the flooding of the river and help in irrigating the surrounding area. Moreover, the kings of the twelfth dynasty used the water of the natural lake of Faiyum in the dry periods. In fact these kings transformed the lake into a huge water reservoir, giving the impression that the lake was an artificial excavation, as reported by Pliny the Elder. He is writing, when discussing the pyramids in his book [8], that two of them are "in the place where Lake Mœris was excavated, an immense artificial piece of water, cited by the Egyptians among their wondrous and memorable works". As the surrounding area changed, since the local branch of the Nile shrank, from about 230 BC, the Bahr Yussef became neglected and Lake Moeris began to dry up, creating a depression in the modern province of Faiyum [7].

As the history is showing, it is undoubtedly necessary for the Toshka Lakes what the researches in Ref.4 are recommending, that is, to undertake measures to preserve these water resources and avoid the problem of the salinity of soil when the lakes dry up. Perhaps, it is too late.

**References**
1. Toshka Lakes, Wikipedia, http://en.wikipedia.org/wiki/Toshka_Lakes
2. In fact, it was with the Mercury missions in the early 1960s, that astronauts began to take pictures of the Earth. The NASA, through its "Gateway to Astronaut Photography of Earth", provides a database with locations, supporting data, and digital images. The images are processed from those coming down from the International Space Station, http://eol.jsc.nasa.gov/. The

"NASA Earth Observatory", created in 1999, has the same task. This is considered the principal source of satellite imagery and other scientific information on climate and environment, among the services provided by NASA, for the general public, http://en.wikipedia.org/wiki/NASA_Earth_Observatory,
http://earthobservatory.nasa.gov/IOTD/view.php?id=1008

3. Gamal M. El-Shabrawy and Henri J. Dumont, The Toshka Lakes, in, The Nile, Monographiae Biologicae, 2009, Volume 89, III, 157-162.
4. M. El Bastawesy, S. Arafat and F. Khalaf, Estimation of water loss from Toshka Lakes using remote sensing and GIS, 10th AGILE International Conference on Geographic Information Science 2007, Aalborg University, Denmark, pp.1-9.
5. Anelia Carolina Sparavigna, Enhancing the Google imagery using a wavelet filter, 8 Sep 2010, Geophysics (physics.geo-ph); Earth and Planetary Astrophysics (astro-ph.EP), arxiv:1009.1590
6. Amelia Carolina Sparavigna, The satellite archaeological survey of Egypt, 31 May 2011, Geophysics (physics.geo-ph), arXiv:1105.6315
7. http://en.wikipedia.org/wiki/Lake_Moeris
8. Pliny, The Natural History, Translated by John Bostock and H.T. Riley, Henry G. Bohm, London, 1857

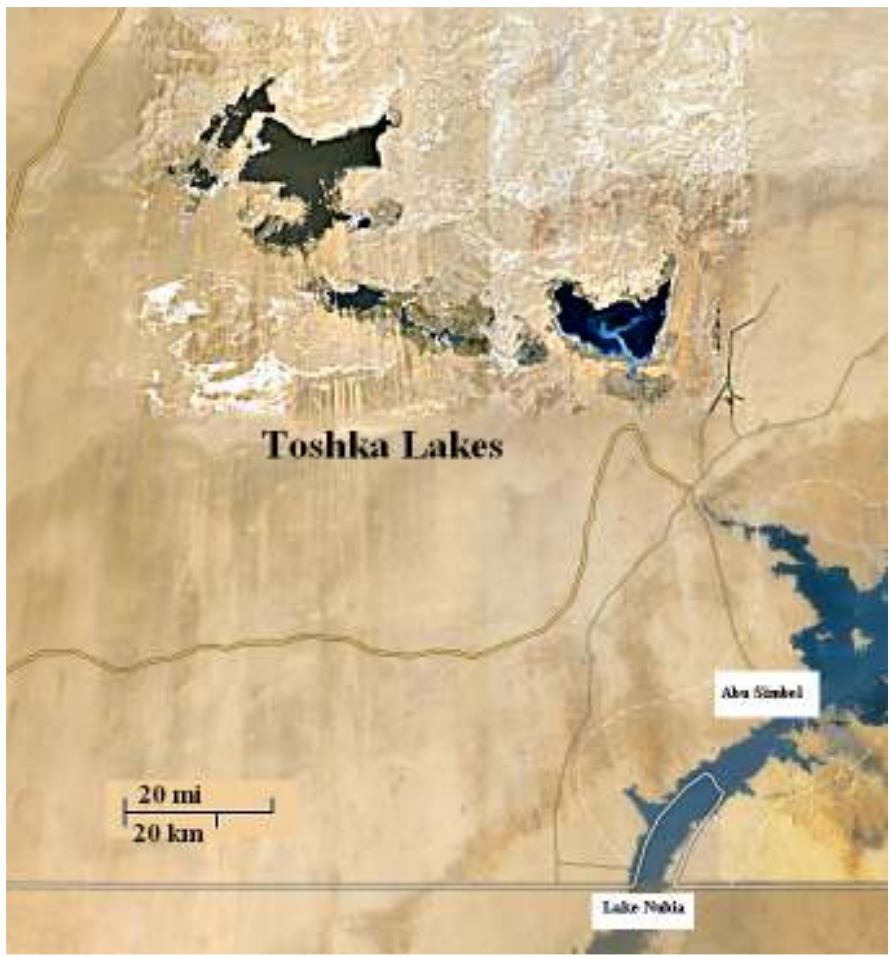

Fig.1 Toshka Lakes are lakes recently formed in the Sahara Desert of Egypt. They are near the Nasser Lake, the huge lake was created by the Aswan High Dam, built in 1964-1968. The water is flowing in the Sadat Canal, pumped in the Toshka Depression, when it is overflowing a specific level in the Lake Nasser. The image has been adapted from the Google Maps.

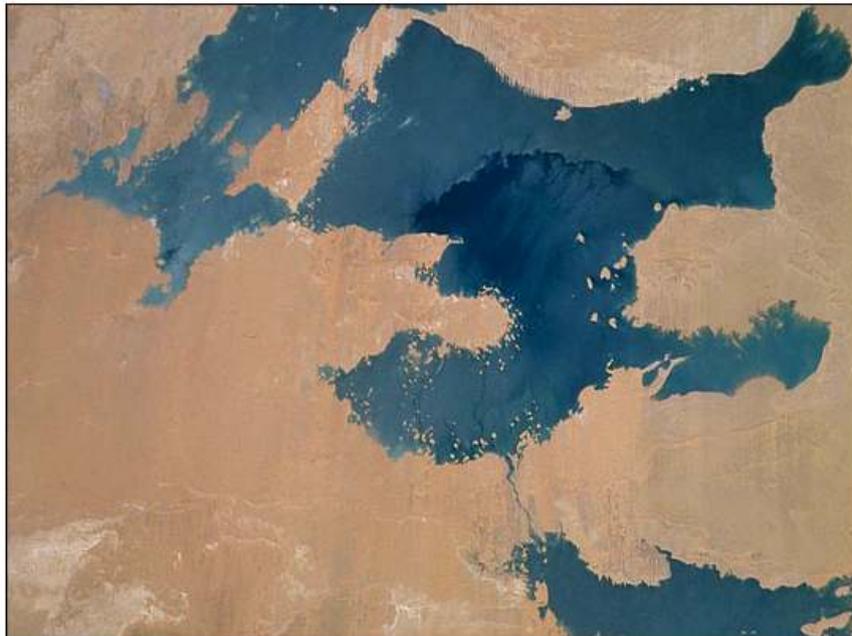

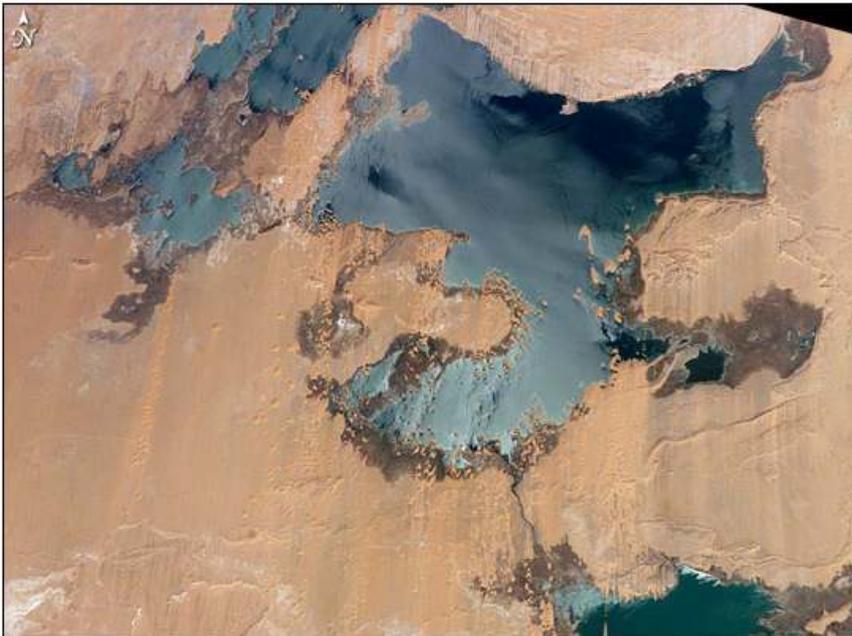

Fig.2 The "NASA Earth Observatory" allows to see the evolution from 2001 to 2005, of the western Toshka Lakes. In the upper image, we have a record of the lakes as soon as they formed. But in 2005, the water level was strongly reduced.

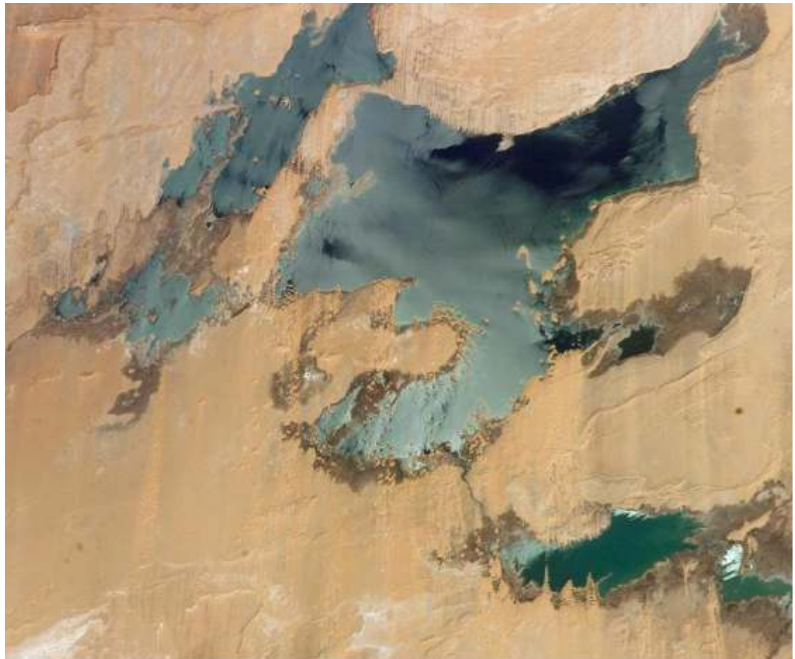

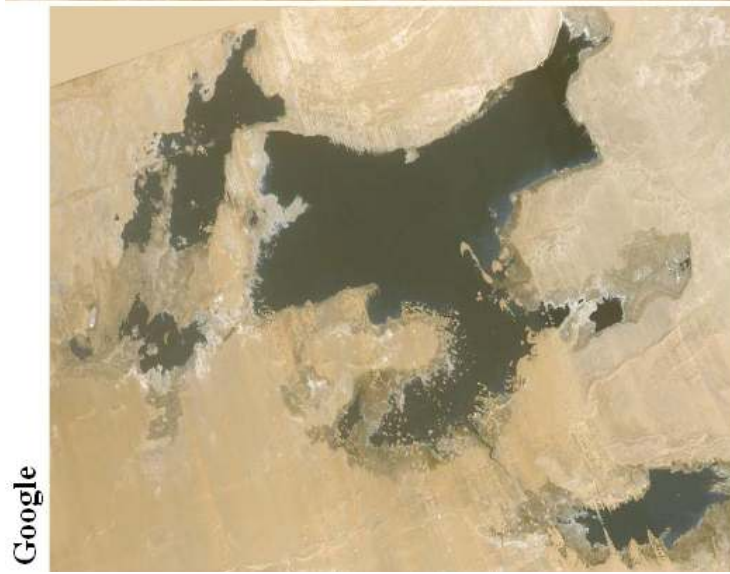

Fig.3 The "NASA Earth Observatory" of the western Toshka Lakes in 2005 (upper panel) is compared with an image from Google Maps (downloaded July 21, 2011). It is not possible for the author to tell when the image of the lower panel had been recorded. But, surely it is more recent. We can see a further reduction of water in the lakes.

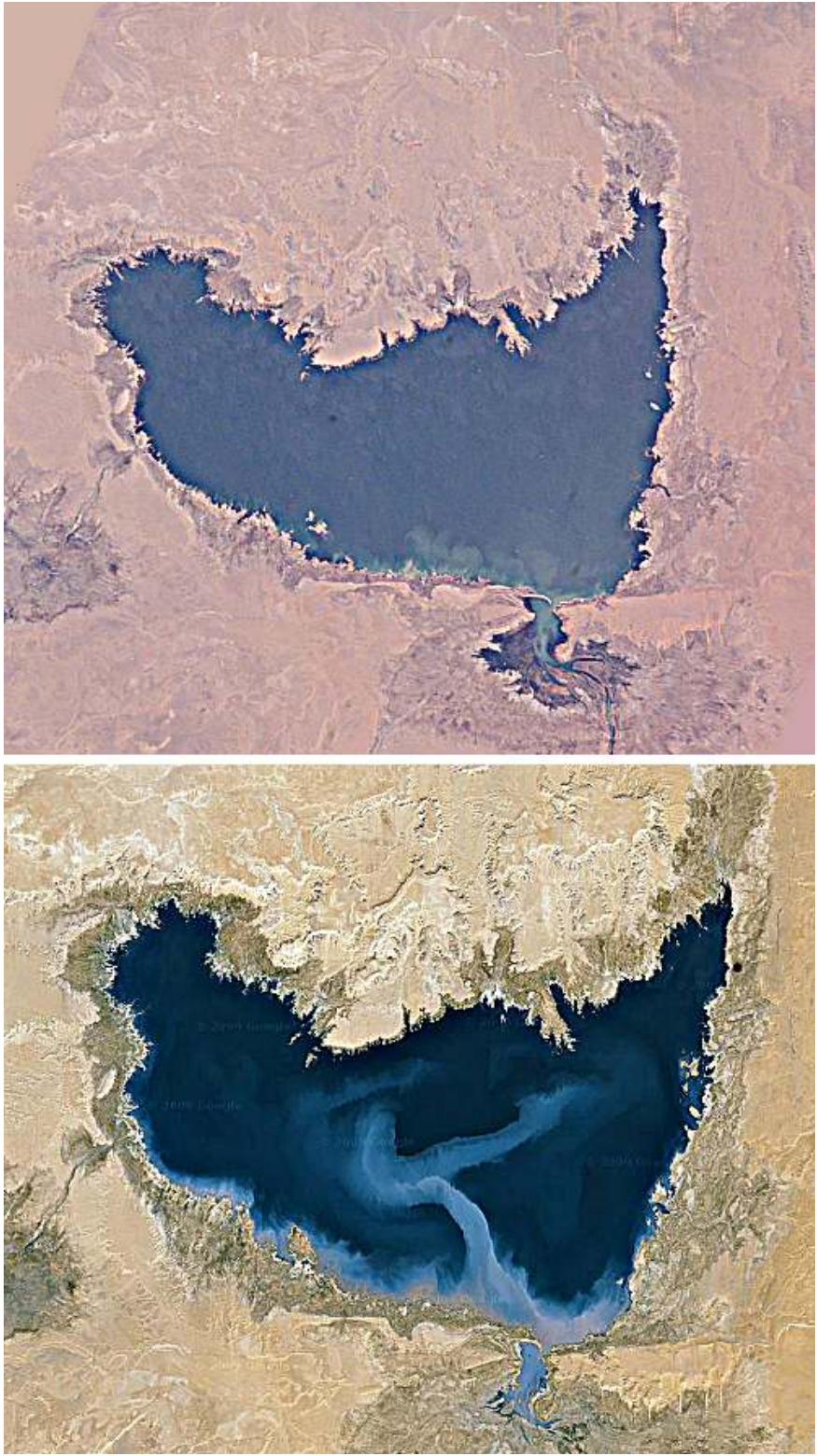

Fig.4 The easternmost Toshka Lake in 2005, as provided by the "Gateway to Astronaut Photography of Earth", is shown in the upper panel. The original image has been enhanced and rotated to compare it with that provided by the Google Maps, displayed in the lower panel. As in the case of Fig.3, it is not possible for the author to tell when the Google image had been recorded, surely after 2005.

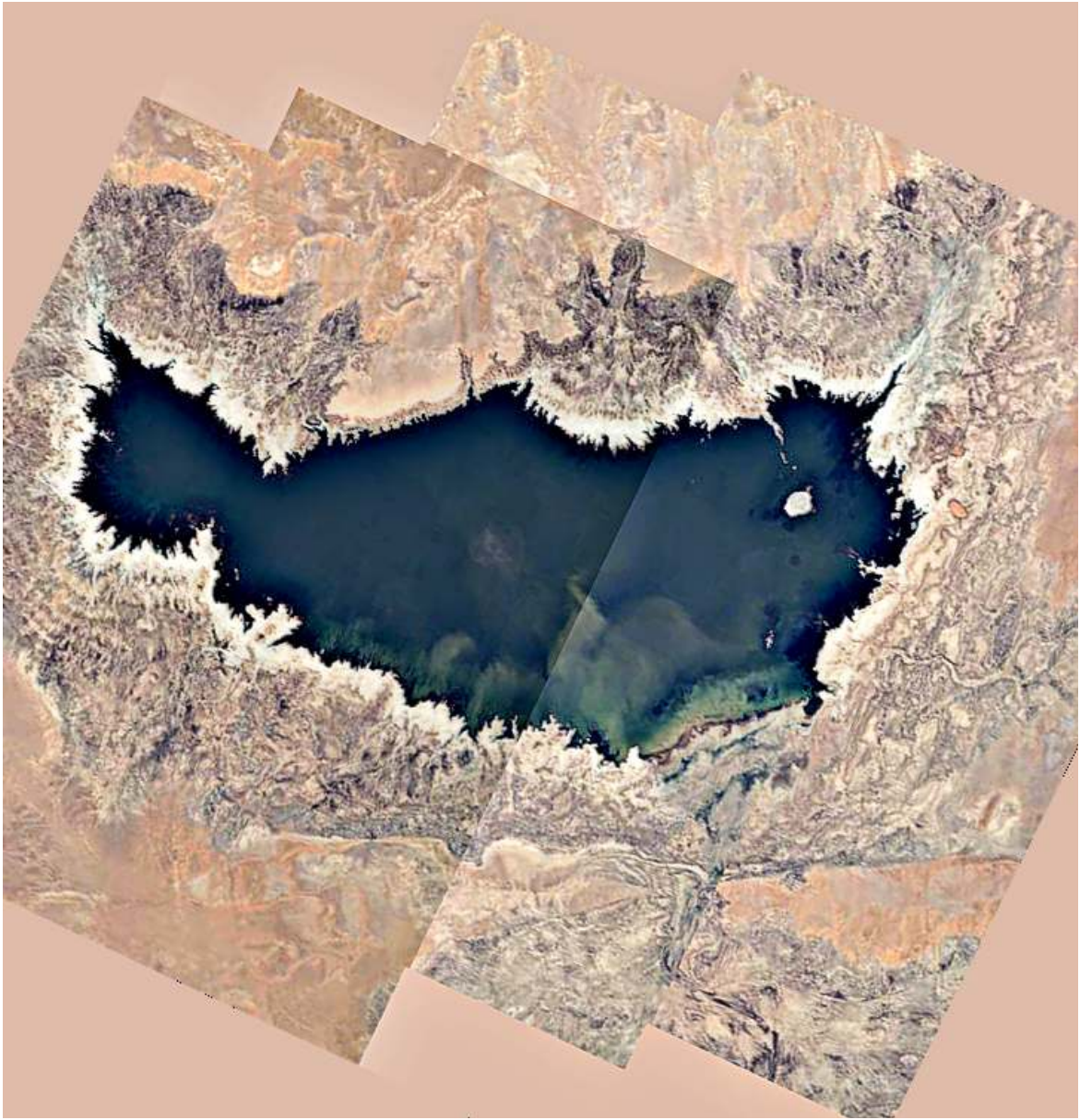

Fig.5 A set of images from the "Gateway to Astronaut Photography of Earth" dated, March 31, 2011, allows to create a composite map of the current aspect of the easternmost lake. This map had been rotated and enhanced as discussed in Ref. 5. The processed image has so many details useful for a comparison with the images in Fig.4.

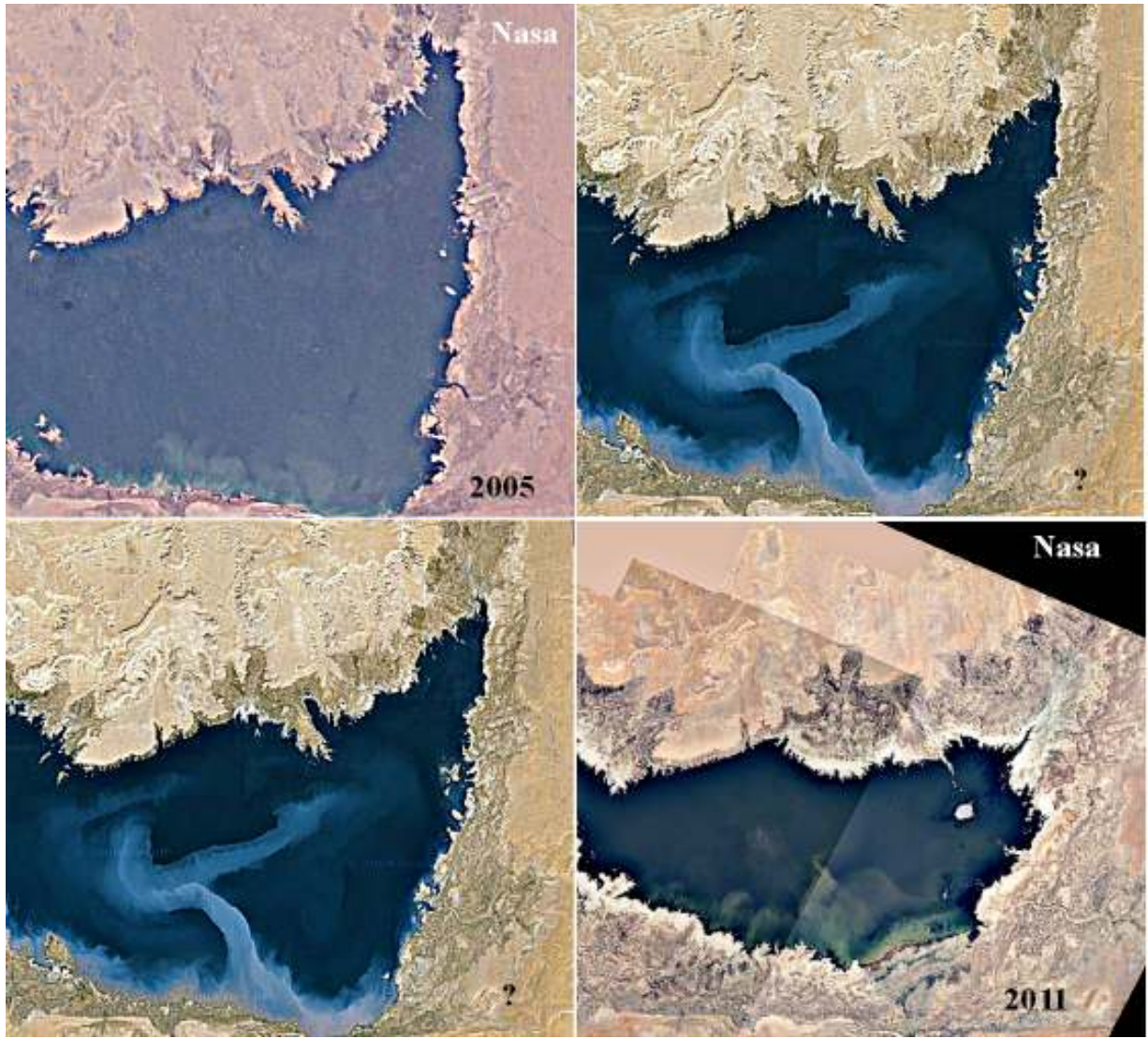

Fig.6. From the image by the "Gateway to Astronaut Photography of Earth" dated 2005 (up-left), to that of March 31, 2011 (down-right), passing through that from Google Maps (up-right/down-left). The two images labelled with a question mark, are the same. The question mark means that the year when it was recorded is unknown to the author. We have the evolution of a part of the easternmost lake that was large in 2005 shrinking between 2005 and 2011, and reduced to one half the original part in 2011. The surface of the lake has dramatically decreased.

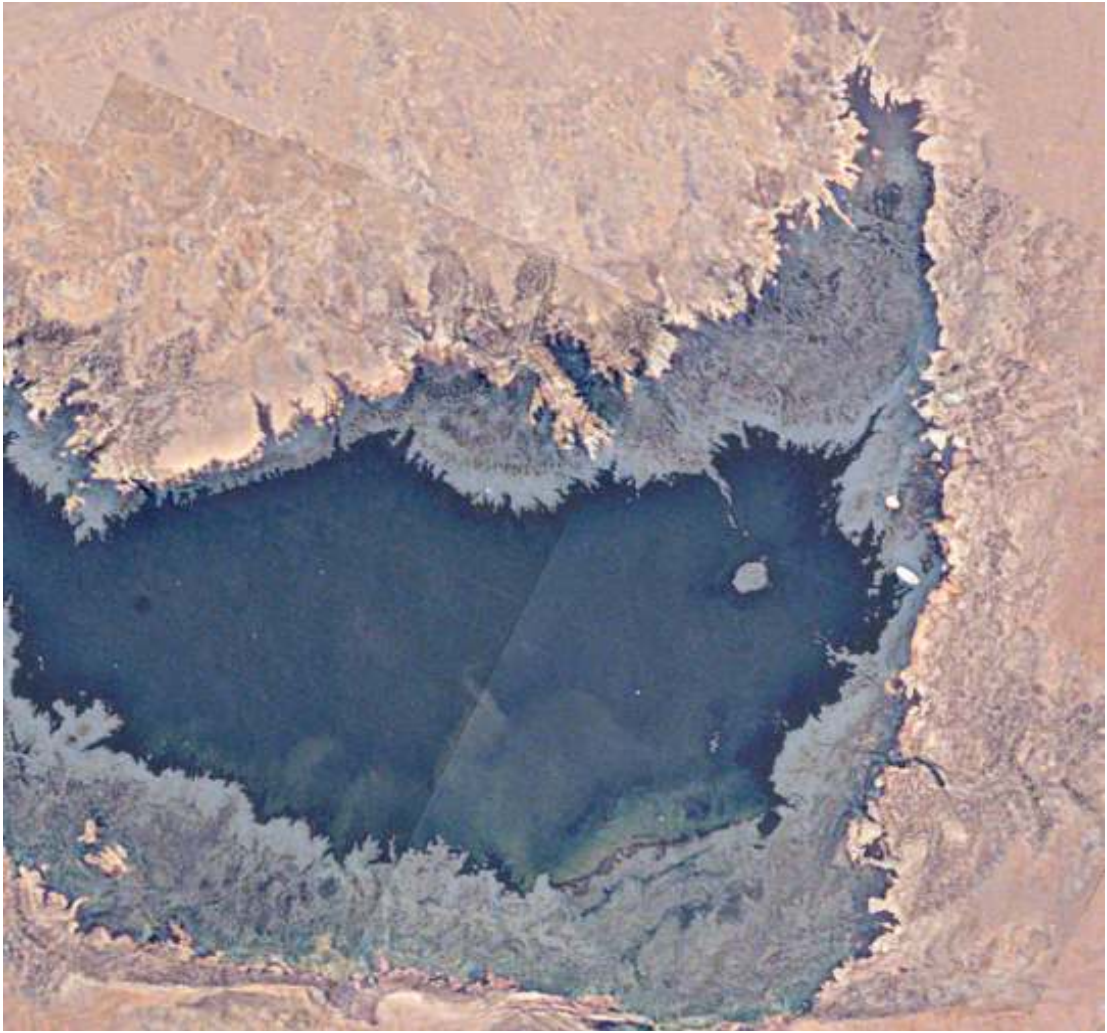

Fig.7. Superposition of the two NASA images, of 2005 and 2011, of Fig.6, proposed to help the reader in the comparison.